# Artificial intelligence based prediction on lung cancer risk factors using deep learning

**Muhammad Sohaib[1], Mary Adewunmi[2]**
[1]College of Artificial Intelligence and Computing, Tianjin University, Tianjin, China
[2]School of Medicine, University of Tasmania, Hobart, Australia

**ABSTRACT**

In this proposed work, we identified the significant research issues on lung cancer risk factors. Capturing and defining symptoms at an early stage is one of the most difficult phases for patients. Based on the history of patients records, we reviewed a number of current research studies on lung cancer and its various stages. We identified that lung cancer is one of the significant research issues in predicting the early stages of cancer disease. This research aimed to develop a model that can detect lung cancer with a remarkably high level of accuracy using the deep learning approach (convolution neural network). This method considers and resolves significant gaps in previous studies. We compare the accuracy levels and loss values of our model with VGG16, InceptionV3, and Resnet50. We found that our model achieved an accuracy of 94% and a minimum loss of 0.1%. Hence physicians can use our convolution neural network models for predicting lung cancer risk factors in the real world. Moreover, this investigation reveals that squamous cell carcinoma, normal, adenocarcinoma, and large cell carcinoma are the most significant risk factors. In addition, the remaining attributes are also crucial for achieving the best performance.

*Corresponding Author:*

Muhammad Sohaib
College of Artificial Intelligence and Computing, Tianjin University
Tianjin, China
Email: muhammad.sohaib@tju.edu.cn

## 1. INTRODUCTION

Lung cancer is one of the world's most serious diseases. Early-stage assessment can save many lives when cancerous tumors are identified early [1]. In addition to examining the patient, blood tests, X-rays, biopsies, and computed tomography (CT) scans can be used to diagnose cancer. The American Cancer Society estimates the lung cancer in the United States for 2023, there are 238,340 new cases of lung cancer and about 127,070 deaths from lung cancer [2]. Lung nodules are small tissue masses that show up as white shadows on CT scans and X-ray imaging. Managing such high-dimensional data create challenges for statistical analysis [3]. Furthermore, when radiographs are less characteristic, health care professional's decision-making ability becomes more difficult in the absence of technical equipment [4]. Lung cancer is one of the most prevalent cancers worldwide, causing 1.76 million deaths each year. Chest CT scans play an essential role in the screening and diagnosing of lung cancer [5]. The randomized controlled trial has shown that low-dose CT screening reduces mortality from lung cancer in high-risk patients, and recent studies have demonstrated the benefit of CT screening in community settings. The widespread adoption of lung cancer screening is expected to benefit millions of people [6]. However, the millions of CT scans obtained from patients create a massive workload for radiologists. In addition, interstate conflicts were recorded. Previous studies have suggested that computer-aided diagnostic systems may improve the detection of pulmonary nodules on CT examination. Artificial intelligence (AI)-based automated CT lung cancer detection is a feasible option that can assist physicians by reducing their workload, improving hospital operational throughput, and providing them with more time to develop high-quality doctor-patient relationships. Numerous studies have indicated that combining a physician/radiologist assessment with the use of an AI or machine learning model improves the detection of lung nodules on CT scans. AI models improved radiologist performance in detecting small lung nodules (less than 5 mm in diameter), which are commonly missed by eye assessment alone. Deep learning models minimize the workload on physicians, reduce fatigue-related errors of judgment, and improve the detection of nodules, which are more likely to be missed in the early stages of lung cancer. This research study provides a unique opportunity to investigate the rigidity of medical deep-learning models and compare the performance of various strategies for processing and classifying large-scale chest CT images [7]. In recent years, many methods have been proposed to diagnose lung cancer, most of which use CT scans and some of which use X-ray images [8].

In this study, we created a deep-learning model to detect lung cancer accurately. We use a publicly available dataset of CT scan images labeled as squamous cell carcinoma, normal, adenocarcinoma, and large cell carcinoma. We compared our model to the VGG16, InceptionV3, and Resnet50 models and discovered that it outperforms by 94%. We also discuss the issue of low accuracy in lung cancer detection due to various factors such as data set type, pre-processing data techniques, deep learning model implementation method, and data set volume used.

## 2. RELATED WORKS

The cancer cells are dangerous and can lead to a high mortality rate. The predictive model provides accurate cancer treatment [9]. Lung cancer is a dangerous type of cancer and is difficult to diagnose. It usually causes death in both sexes, so it is essential to check the nodules immediately and correctly [10]. The images of chest CT scans play an important role in screening and diagnosing lung cancer [5]. The early detection of cancer plays an important role in saving lives. Early detection gives the patient a better chance of recovery. Technology plays a major role in the effective detection of cancer. Several methods have been proposed to diagnose lung cancer at an early stage [8]. The research aims to compare the approach and performance of award-winning algorithms that have proposed and developed reproductive machine learning modules to detect lung cancer. Kadir and Gleeson [11] suggested that machine learning-based lung cancer diagnostic models have been proposed to assist physicians in managing randomized or screen-detected indeterminate pulmonary nodules. Such systems can reduce diversity in nodule classification, improve decision making, and ultimately reduce the number of benign nodules that unnecessarily bind or act on them. Obulesu et al. [12] emphasizes that cancer is a worldwide health problem with increased mortality in recent years. Therefore, fortunately, progressive machine learning methods can distinguish lung cancer patients from healthy individuals, which is of great concern. Chaturvedi et al. [13] emphasize that lung cancer has once again emerged as the most common disease in humankind. Early detection gives the patient a better chance of recovery. The predictive model plays a significant role in detecting cancer, but the machine learning model was not highly accurate to give 100% assurance. Radhika et al. [14] stated that the growth of cancer cells in the lungs is called lung cancer.

The increasing cancer rate has resulted in an increase in the death rate for both men and women. Research by Vikas and Kaur [15] claim that lung cancer is the most common disease in the world. Machine learning algorithms make it easier to diagnose and diagnose lung cancer. For segmentation of images, convolutional neural networks (CNN) is modified and different architectures are formed as eNet [16], U-Net [17], DenseNet [18], SegNet [19], FCN [20], DilatedNet [21], PixelNet [22], ERFNet [23], DeconvNet [24], and ICNet [25]. The Kaggle data science bowl competition was held to increase the algorithm's accuracy in classifying and detecting if the lungs nodules in CT scan images are malignant. In stage 1, a large training dataset of 1,397 patients (362 with lung cancer and 1,035 without) was provided, along with an initial validation set of 198 patients. This validation set was used to generate the public stage 1 leader-board, which competitors could use to evaluate their performance. In stage 2, a previously unseen dataset of 506 patients was made available for 7 days to judge the final competition results. This two-stage approach was used to avoid competitors inferring test set labels from a large number of entries. In contrast to the LungX competition, competitors here had to create a fully automated pipeline that took in a CT image and output a likelihood of cancer [26].

Research by Göltepe [27] addressed the fact that late diagnosis of lung cancer causes deaths. An early diagnosis dramatically increases the chances of successful lung cancer treatment. Machine-learning techniques are developing rapidly and are increasingly being used by medical professionals to classify and diagnose early-stage diseases. Sajad et al. [28] emphasizes that lung cancer is the leading cause of cancer-related deaths. X-ray images of the lung can reveal abnormal masses or nodules, but accurate cancer diagnosis in the laboratory can take a long time. This study uses deep learning techniques to detect cancer at an earlier stage based on the case histories of previous patients.

## 3. DESCRIPTION OF DATASET

The data statistics in Figure 1 show that 613 images belong to 4 classes: squamous cell carcinoma, normal, adenocarcinoma, and large cell carcinoma. The categories of data are based on the history of patients who are accessed from the Kaggle [29]. In this research study, we used the 80:20 splitting data ratio, which signified that 80% are used to train the predictive model, whereas 20% of data are used to test and validate the accuracy level at different algorithms.

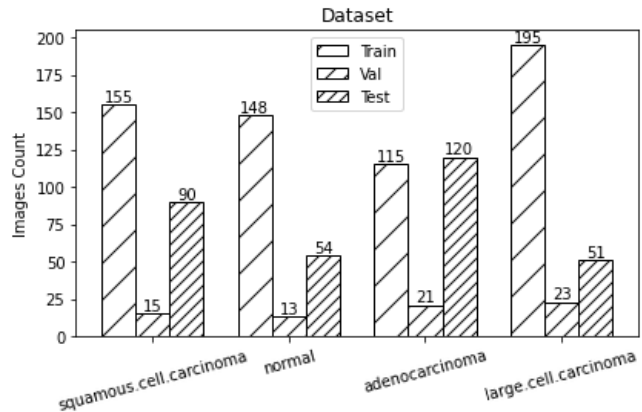

Figure 1. Description of dataset

## 4. RESEARCH DESIGN AND METHOD

In this research article, we develop the CNN algorithm to predict the categories of chest CT scan image classes such as squamous cell carcinoma, normal, adenocarcinoma, and large cell carcinoma. The proposed system was implemented using the Jupyter editor and the Python 3.8 programming language via the anaconda distribution. The model was built using the TensorFlow framework and Google Colab. The proposed architecture is presented in Figure 2.

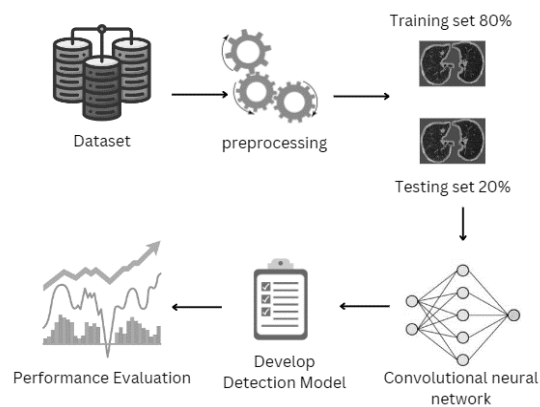

Figure 2. Proposed architecture

In comparison to other neural network models, the CNN family of neural networks performs exceptionally well in applications involving image, speech, and audio processing. In neural networks, each edge has a weight that is attached to it. It is made up of layered sets, where a layer is a set of nodes. Both an input layer and an output layer are part of it. The hidden layer is located between these two layers and varies depending on the type of network. When an input signal is applied, each node generates output that serves as input to the hidden layers. The output of a given node is calculated by applying an activation function (ψ) to the inputs of the previous layers' nodes. The process described above is known as inference. The algorithm is explained as:

CNNs are a type of neural network architecture designed for image and signal processing tasks. The structure of a CNN can be explained using several variables (Table 1). The total number of layers in the network is denoted as L (total number of layers contained in the neural network), while the $k^{th}$ layer is represented by $l^{(k)}$ ($k^{th}$ layer). Each layer $l^{(k)}$ contains a certain number of nodes, denoted by $m^{(k)}$ (nodes contained in $l^{(k)}$), with each node $i$ in layer $l^{(k)}$ represented as $l_i^{(k)}$ (node $i$ in layer $l^{(k)}$). The output vector representing the outputs of the nodes in layer $l_{(k)}$ is denoted as $O^{(k)}$ (the output vector representing the outputs of the nodes in $l^{(k)}$), while the output of a specific node $l_i^{(k)}$ is represented as $O_i^{(k)}$ (layer $l_i^{(k)}$ output). The matrix's weight that connects nodes in layer $l^{(k-1)}$ to nodes in layer $l^{(k)}$ is denoted by $w^{(k)}$ (matrix's weight that connects nodes in layer $l^{(k-1)}$ to nodes in layer $l^{(k)}$), with the weight connecting nodes $l_i^{(k-1)}$ and $l_j^{(k)}$ represented as $w_{i,j}^{(k)}$ (weight connecting nodes $l_i^{(k-1)}$ and $l_j^{(k)}$). The bias for layer $l^{(k)}$ is denoted as $b^{(k)}$ (bias for $l^{(k)}$). The function used to determine the outputs $O^{(k)}$ and $O^{(k-1)}$ is denoted as $\psi^{(k)}$ (function to determine the outputs $O^{(k)}$ and $O^{(k-1)}$), while the activation function used in each layer is denoted as sigma. In terms of inputs and outputs, X represents every input data in the dataset, while Y represents every provided output of the dataset. The approximation of all outputs given all inputs is denoted as Ŷ. For a specific nth data input, it is represented as $x_n$, while the specific nth output layer is represented as $y_n$. The approximation of output $y_n$ given input $x_n$ is represented as $ŷ_n$ (approximation of output $y_n$ given input $x_n$).

Table 1. Overview of studies using deep learning for lung cancer detection

| Study | Method | Libraries/language/software used for simulation and implementation | Accuracy (%) |
| --- | --- | --- | --- |
| [30] | DenseNet | K means/Gaussian naïve bayes/support vector machines | 92 |
| [31] | DCNN | Python/Pytorch/Adam optimizer | 63 |
| [32] | Deep residual learning | XGBoost /random forest | 84 |
| [33] | Gould model | Bootstrap method | 81 |
| Our model | CNN developed model | Python/TensorFlow/Google Colab | 94 |

## 5. RESULTS AND DISCUSSION

The research presented a model that achieved an exceptional accuracy rate of 94% after 32 iterations, which is a notable accomplishment. Figure 3 provides a clear visualization of the model's various layers, serving as a useful summary of the model's architecture. Overall, this study contributes to the field of machine learning by presenting a high-performing model and detailing its layer structure.

```
Layer (type)                    Output Shape              Param #
=================================================================
conv2d (Conv2D)                 (None, 348, 348, 32)      896

conv2d_1 (Conv2D)               (None, 346, 346, 32)      9248

max_pooling2d (MaxPooling2D     (None, 173, 173, 32)      0
)

conv2d_2 (Conv2D)               (None, 171, 171, 64)      18496

max_pooling2d_1 (MaxPooling     (None, 85, 85, 64)        0
2D)

conv2d_3 (Conv2D)               (None, 83, 83, 128)       73856

max_pooling2d_2 (MaxPooling     (None, 41, 41, 128)       0
2D)

dropout (Dropout)               (None, 41, 41, 128)       0

flatten (Flatten)               (None, 215168)            0

dense (Dense)                   (None, 64)                13770816

dropout_1 (Dropout)             (None, 64)                0

dense_1 (Dense)                 (None, 4)                 260

=================================================================
Total params: 13,873,572
Trainable params: 13,873,572
Non-trainable params: 0
_________________________________________________________________
```

Figure 3. Extract of model training

We use the skip connections to add output from the previous layer to the next layer. This will help reduce the vanishing gradient problem. During the computation, we discovered that the accuracy of the developed model after epoch 00032 did not improve from 0.94286, whereas the loss value is 1.3498, which is highly significant for this research study (Figure 4).

After building the convolution neural network model on 613 images from four classes (squamous cell carcinoma, normal, adenocarcinoma, and large cell carcinoma), we discovered that our predictive model passes through an accuracy level 0.94286, whereas the loss value is 1.3498. We evaluate the performance of the proposed model with three existing algorithms i.e., VGG16, InceptionV3, and Resnet50. We discovered that after 32 epoch, the VGG16 model accuracy is 0.8857 whereas the loss value is 0.8196. The InceptionV3 model accuracy is 0.9142, and the loss value is 0.4118. Resnet50 model has a value accuracy 0.9142 with the loss value 0.2457. The accuracy of models as shown in Figure 5.

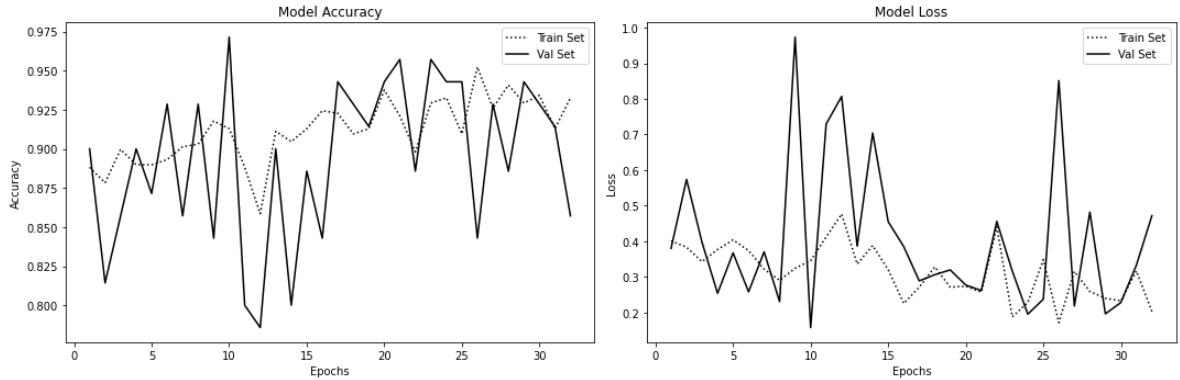

Figure 4. Loss and accuracy values of the developed model

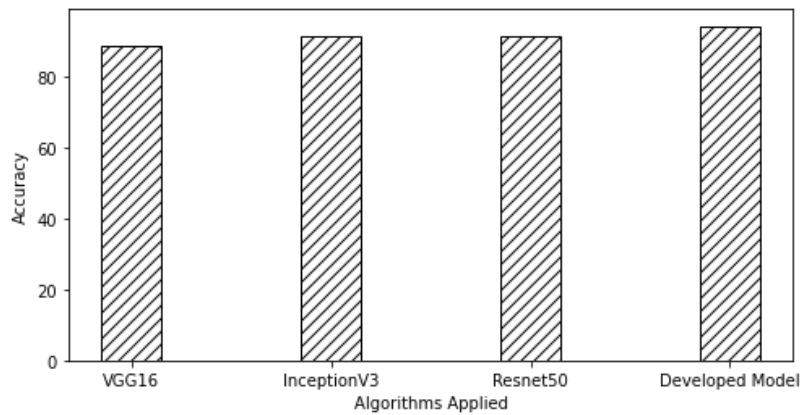

Figure 5. Statistical analysis of models

To evaluate our model effectiveness, we compared it to several existing models, as outlined in Table 1. After conducting a thorough data analysis, we determined that our proposed CNN model provided more accurate and efficient predictions for lung cancer diagnosis. Overall, this research demonstrates the potential for advanced machine learning techniques to improve medical diagnoses and outcomes.

## 6.   CONCLUSION

This research study is based on the recurrent issue of lung cancer detection. We emphasized the value of early cancer detection in terms of saving lives. Early detection increases the chances of recovery for the patient. Technology is critical for early cancer detection. Due to the large amount of data and blurred boundaries in CT images, tumor differentiation and classification are difficult. This study presented an automated lung cancer detection method to improve accuracy and yield while decreasing diagnostic time. We developed a CNN, train on 613 images belonging to 4 classes such as squamous cell carcinoma, normal, adenocarcinoma, and large cell carcinoma. The evaluation showed that our model achieved an accuracy of 94% and a minimum loss of 0.1, which is a relative improvement as compared to existing systems. The proposed model successfully produce correct results, which reduce human error mistakes in the diagnosis process and decrease the cost of lung cancer diagnosis. This study's primary weakness was its reliance on a secondary dataset, like Kaggle. In the future, other primary data will be used to increase the model's accuracy. We also intend to create a desktop application tool so that physician can use the model. This tool will help in diagnosis of the lung cancer by just feeding the CT scan images.

## BIOGRAPHIES OF AUTHORS

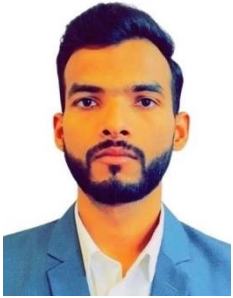
**Muhammad Sohaib** is an accomplished AI research scientist at Caresai, currently pursuing a master's degree in Electronic Information at Tianjin University, China. With a bachelor's degree in Computer Science from GC University Lahore. He has honed his skills in building AI and ML applications across multiple platforms. His research interests revolve around the fascinating fields of artificial intelligence, machine learning, deep learning, and natural language processing. He is widely recognized for his ability to creatively solve complex problems, a skill that has enabled him to deliver innovative solutions in the field of AI technology. He can be contacted at email: muhammad.sohaib@tju.edu.cn.

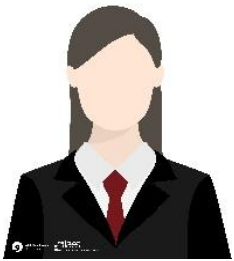
**Mary Adewunmi** is a PhD Student in the School of Medicine, at the University of Tasmania (UTAS), Hobart, Australia. She is a Computer Science graduate of Bowen University, Iwo, Osun State, and a master's graduate of Lagos State University, Ojo, Nigeria. She works full-time as a medical AI research officer at UTAS, lead machine learning engineer and group head of Caresai, as a freelance lead machine learning engineer at Omdena. Her career goal is to speed up disease diagnosis and treatment using AI methods. She can be contacted at email: mary.adewunmi@utas.edu.au.